\documentclass[superscriptaddress,secnumarabic,amssymb,amsmath,12pt,nobibnotes,aps,prd,showkeys,nofootinbib]{revtex4}
\usepackage{amsmath}

\newtheorem{remark}{Remark}[section]
\usepackage{graphicx}
\usepackage{epsf}
\usepackage{bm}
\usepackage{amsmath}
\usepackage{amsfonts}
\usepackage{amssymb}
\usepackage{epstopdf}
\usepackage{natbib}
\usepackage{color}

\begin{document}
\tolerance=5000
\noindent
%--------------------------------------------------------------------
\title{The Peebles - Vilenkin quintessential inflation model revisited}

\author{Jaume Haro}
\email{jaime.haro@upc.edu}
\affiliation{Departament de Matem\`atiques, Universitat Polit\`ecnica de Catalunya, Diagonal 647, 08028 Barcelona, Spain}

\author{Jaume Amor\'{o}s}
\email{jaume.amoros@upc.edu}
\affiliation{Departament de Matem\`atiques, Universitat Polit\`ecnica de Catalunya, Diagonal 647, 08028 Barcelona, Spain}

\author{Supriya Pan}
\email{supriya.maths@presiuniv.ac.in}
\affiliation{Department of Mathematics, Presidency University, 86/1 College Street, Kolkata 700073, India.}

\begin{abstract}

We review the well-known Peebles-Vilenkin (PV) quintessential inflation model and discuss its possible improvements in agreement with the recent observations.   
The improved PV model  depends only on two parameters:
the inflaton mass $m$,  and another smaller mass $M$; where the latter has to be chosen in order to undertake that, at present time,  the dark energy density of the universe is approximately about 70\% of the total energy budget of the universe. The value of the inflaton mass $m$ is calculated using the observational value of the power spectrum of the scalar perturbations, and the value of mass $M$, which depends on the reheating temperature, is calculated by  solving  the corresponding dynamical system whose initial conditions are taken at the matter-radiation equality and are obtained from three observational data: the red shift at the matter-radiation equality, the ratio of the matter energy density to the critical one at the present time and the current value of the Hubble parameter.
\end{abstract}

\vspace{0.5cm}

%-------------------------------------------------------------------------
\maketitle
%------------------------------------------------------------------------
\thispagestyle{empty}
\tableofcontents

\section{Introduction}

The understanding of physical cosmology at the fundamental level is quite cloudy, especially when a unified picture of the universe's evolution is searched for. The  
inflation \cite{Starobinsky,guth,linde,Linde:1982uu,Burd:1988ss,Barrow:1990vx,Barrow:1994nt} $-$ a rapid accelerating stage of the early universe $-$ and quintessence \cite{Caldwell:1997ii,Carroll:1998zi,Steinhardt:1999nw,Chiba:1999wt,Sahni:1999qe,Barreiro:1999zs,UrenaLopez:2000aj,Yang:2018xah} $-$ another accelerating stage of the late universe $-$ are two significant phases of the universe's evolution which are probably the main focus areas of the scientific community at present.  Since inflation and quintessence present two different stages of the universe evolution, typically, they are considered as the effects of two different exotic sources. The scalar field theory has been found to be an excellent candidate for the inflationary scenario and for quintessence phase, another weakly interacting scalar field is usually considered. The common feature of these scenarios is that both of them should have an accelerating phase, but at the end  are taking two different fields one for the inflation and the other for the quintessence.  This disparity naturally raised a question to Peebles and Vilenkin (PV) and as a consequence they introduced, for the first time, a potential for the scalar field that allows inflation for early universe and quintessence for the current universe. This model by Peebles and Vilenkin is widely known as the  ``{\it quintessential inflation}'' \cite{pv}. The introduction of the quintessential inflationary model also triggered other investigators \cite{Giovannini:1999bh,dimopoulos1,Giovannini:2003jw,hossain1,hossain2,hossain3,hap1,deHaro:2016hsh,deHaro:2016ftq,hap,deHaro:2017nui,Geng:2017mic,AresteSalo:2017lkv,Haro:2015ljc}.

Without any doubt, the ``{\it quintessential inflation}'' model by Peebles and Vilenkin is really elegant by its construction. The model is also very simple because it depends only on two parameters, one which characterizes the inflation and the other characterizes the quintessence phase. The potential has two pieces,  one for early inflation and the other for quintessence. The inflationary piece is quartic in the PV model \cite{pv} and it is matched abruptly with a inverse power law potential (i.e., the quintessence potential) which is responsible to drive the current cosmic acceleration. Due to this abrupt matching,  a phase transition from inflation to a kination regime occurs \cite{Joyce}, where the adiabatic evolution of the universe is broken, and thus, the particles are produced following a reheating mechanism, such as, instant preheating or gravitational particle production of heavy massive or massless particles.

Due to simplicity and potentiality, the original PV model naturally gained a massive attention in the cosmological community. However,  from the recent observational predictions, the model has been dignosed with some limitations. In particular, it was found that the quartic inflationary piece of this model leads to theoretical values of the spectral index ($n_s$) and the ratio of tensor to scalar perturbations ($r$) which do not enter into the corresponding two-dimensional marginalized joint contour at $95\%$ Confidence Level  \cite{Planck}. While interestingly, if the quartic part of the inflationary potential is turned into quadratic one,  then the aforementioned problem does not encounter.  That means if the quartic potential is changed by a quadratic one, then for the typical number of $e$-folds in the quintessential inflation, i.e., between $60$ and $75$  (see for instance \cite{deHaro:2016ftq,deHaro:2017nui}), the theoretical values provided by this potential enters in this contour \cite{hap}. Thus, the parameter characterizing this piece of the  potential is the mass of the inflaton field, which is determined from the value of the power spectrum of scalar perturbations when the pivot scale leaves the Hubble radius.
Moreover, the reheating mechanism in \cite{pv} is the gravitational particle production of massless particles which gives a reheating temperature of the order of $1$ TeV, which according to the observational predictions seems to be not enough to solve the overproduction of Gravitational Waves (GW) and as a result this could affect the success of the Big Bang Nucleosynthesis process.

Thus, looking into the observational limitations of the PV model, in the present article we investigate the consequences of the replacement of the original quartic piece of the inflationary potential by the quadratic one and to study the evolution of the model up to the present epoch, which is related to the numerical value of the parameter $M$ that characterizes the  quintessence piece of the potential.
In order to perform the  calculations from both analytical and  numerical grounds, we need three observational parameters, namely the redshift at the
matter-radiation equality,  the ratio of the matter energy density to the critical one at the present time and the current value of the Hubble parameter, 
which in the present work are chosen to be the central values of those parameters, obtained by the $\Lambda$CDM based Planck's estimation \cite{planck}.

The paper is organized as follows: In Section \ref{sec-2} we present our improved version of the original Peebles-Vilenkin quintessential inflation model, and its  dynamics, i.e., the evolution of the universe,  is studied  from the beginning of inflation up to the end of the kination epoch.
Section \ref{sec-kin-to-mr} is devoted to the study of the evolution of the universe from the end of kination to 
the matter-radiation equality for two different situations, namely, the massive particles produced at the end of inflation decay
into lighter ones -in order to produce a relativistic plasma in thermal equilibrium- before or after the end of the kination phase.
In Section \ref{sec-mat-rad-equality} we study the evolution of the universe from the matter-radiation equality to the present where mostly we perform numerical calculations. 
The dynamics given by a quintessence exponential potential is studied in Section \ref{sec-exponential}. Finally, we conclude the present work in Section \ref{sec-summary} with all the findings in brief. 
The units used throughout the paper are, $\hbar=c=1$, and we denote  the reduced Planck's mass by 
$M_{pl}\equiv \frac{1}{\sqrt{8\pi G}}\cong 2.44\times 10^{18}$ GeV.

\section{The original model}
\label{sec-2}

To explain the evolution of the universe unifying the early inflation with the current cosmic acceleration, Peebles and Vilenkin  proposed  a  simple and elegant model   based on the following potential \cite{pv}
\begin{eqnarray}
V(\varphi)=\left\{\begin{array}{ccc}
\lambda(\varphi^4 + M^4) & \mbox{for} & \varphi\leq 0\\
\lambda\frac{M^8}{\varphi^4+M^4} &\mbox{for} & \varphi\geq 0,\end{array}
\right.
\end{eqnarray}
where $\lambda$ and $M\ll M_{pl}$ are two free parameters of the model. The parameter $\lambda$ is dimensionless that must be adjusted so that the theoretical 
values provided by the model coincide with the observed ones. The model contains an abrupt phase transition from inflation to kination \cite{Joyce} at $\varphi\cong 0$. This phase transition is needed, because the adiabatic evolution is broken at this stage, thus, the phase transition enables to create an enough number of gravitational particles which consequently reheats the universe. These created particles after their decay as well as interacting with different fields, form a relativistic fluid in thermal equilibrium whose 
energy density eventually dominates over the one of the inflation up to the present time. At present time, the energy density of the inflaton field becomes dominant once again in order to depict the current cosmic acceleration.

The first observational limitation of the model comes from the quartic inflationary potential as follows. For a number of $e$-folds in the range $60$ and $75$,  which is usual in quintessential inflation, the spectral index, namely  $n_s$, and a ratio of tensor to scalar perturbations, namely  $r$, do not enter in the marginalized joint confidence contour in the plane $(n_s, r)$  at $2\sigma$ C.L. However, the values provided by a quadratic potential enter in this contour. Thus, it seems that the inflationary piece of the model might  be changed, for example from quartic potential to  quadratic potential, in order to match with the recent observational data \cite{Planck,planck, planck18,planck18a}.

Another observational limitation of the model comes from the reheating mechanism. In the original PV model, the authors consider a reheating mechanism via gravitational particle production of massless particles that allows one to obtain a reheating temperature in the TeV regime. This reheating temperature is compatible with some of the usual  Big Bang Nucleosynthesis (BBN) bounds, but it cannot prevent the overproduction of the Gravitational Waves (GWs). To overpass this problem one can consider other kind of reheating mechanisms such as the instant preheating \cite{fkl0, fkl} or the gravitational production of heavy massive particles \cite{haro18, hashiba}.  
Effectively,  due to the  phase transition from inflation to kination, there is an overproduction of GWs (see for instance Section 5 of \cite{hap19}). Then, in order that this overproduction does not alter the BBN success, at the reheating time, the ratio of the energy density of GWs to the energy density of the produced particles has to be less than $10^{-2}$ \cite{pv}. As was shown in  \cite{Dimopoulos}, this bound is satisfied when the reheating occurs via instant preheating, and in the case that the reheating is via gravitational production of superheavy particles, the bound is only satisfied when its decay in lighter particles is after the end of the kination period (see for instance \cite{haro18}).

Thus, based on the above issues, in the present work we  choose the following  family  of models improving the  old version of the quintessential inflation \cite{pv}: 
\begin{eqnarray} \label{improved}
V_{\alpha}(\varphi)=\left\{\begin{array}{ccc}
\frac{m^2}{2}(\varphi^2-M_{pl}^2 + M^2) & \mbox{for} & \varphi\leq -M_{pl}\\
\frac{m^2}{2}\frac{M^{\alpha+2}}{(\varphi+M_{pl})^{\alpha}+M^{\alpha}} &\mbox{for} & \varphi\geq -M_{pl},\end{array}
\right.
\end{eqnarray}
where  $\alpha>0$ is a positive dimensionless parameter which parametrizes the family and  $m\cong 5\times 10^{-6} M_{pl}$ is the mass of the inflaton field
which is calculated from  the observational estimation of power spectrum of the scalar perturbations, ${\mathcal P}_{\zeta}\cong \frac{H_*^2}{8\pi^2M_{pl}^2\epsilon_*}\sim 2\times 10^{-9}$ \cite{btw} (see also \cite{haro18a} for a detailed derivation of the result where $\epsilon$ is the main slow-roll parameter and the star means that the quantities are evaluated when the pivot scale leaves the Hubble radius).  
Note also that in our improved version the phase transition is more abrupt compared to the original one, because for the
potential (\ref{improved}), the first derivative is discontinuous at the matching point, while in the original PV model, the fourth derivate is discontinuous at the matching point. This fact is very important in order to obtain an enough amount of massive particles leading to a viable reheating temperature via production of heavy massive particles which do not happen in the original PV model where the reheating via massive particle production leads  to an abnormally small reheating temperature \cite{hap1}.

On the other hand, 
the kination phase starts at $\varphi_{kin}\cong -M_{pl}$,  so assuming, as usual, that there is no substantial drop of energy between the end of inflation and the beginning of kination, one will have 
$\dot{\varphi}_{kin}=\sqrt{6}H_{kin}M_{pl}\cong 7\times 10^{-6} M_{pl}^2$, 
where the overdot represents the differentiation with respect to the cosmic time and $H_{kin}\cong H_{end}\cong \sqrt{\frac{V(\varphi_{end})}{3M_{pl}^2}}=\sqrt{\frac{1+\sqrt{3}}{6}}m\cong 3\times 10^{-6} M_{pl}$, because at the end of inflation, the scalar field reduces to,  
$\varphi_{end}=-\sqrt{2+\sqrt{3}}M_{pl}$.  Since during kination,  $a\propto t^{1/3}\Longrightarrow H=\frac{1}{3t}$, using the Friedmann equation, the dynamics in this regime will be
\begin{eqnarray}
\frac{\dot{\varphi}^2}{2}=\frac{M_{pl}^2}{3t^2}\Longrightarrow \dot{\varphi}=\sqrt{\frac{2}{3}}\frac{M_{pl}}{t}\Longrightarrow 
\varphi(t)=\varphi_{kin}+\sqrt{\frac{2}{3}}M_{pl}\ln \left( \frac{t}{t_{kin}} \right).\end{eqnarray}

\

\begin{figure}
%\begin{center}
\includegraphics[width=0.6\textwidth]{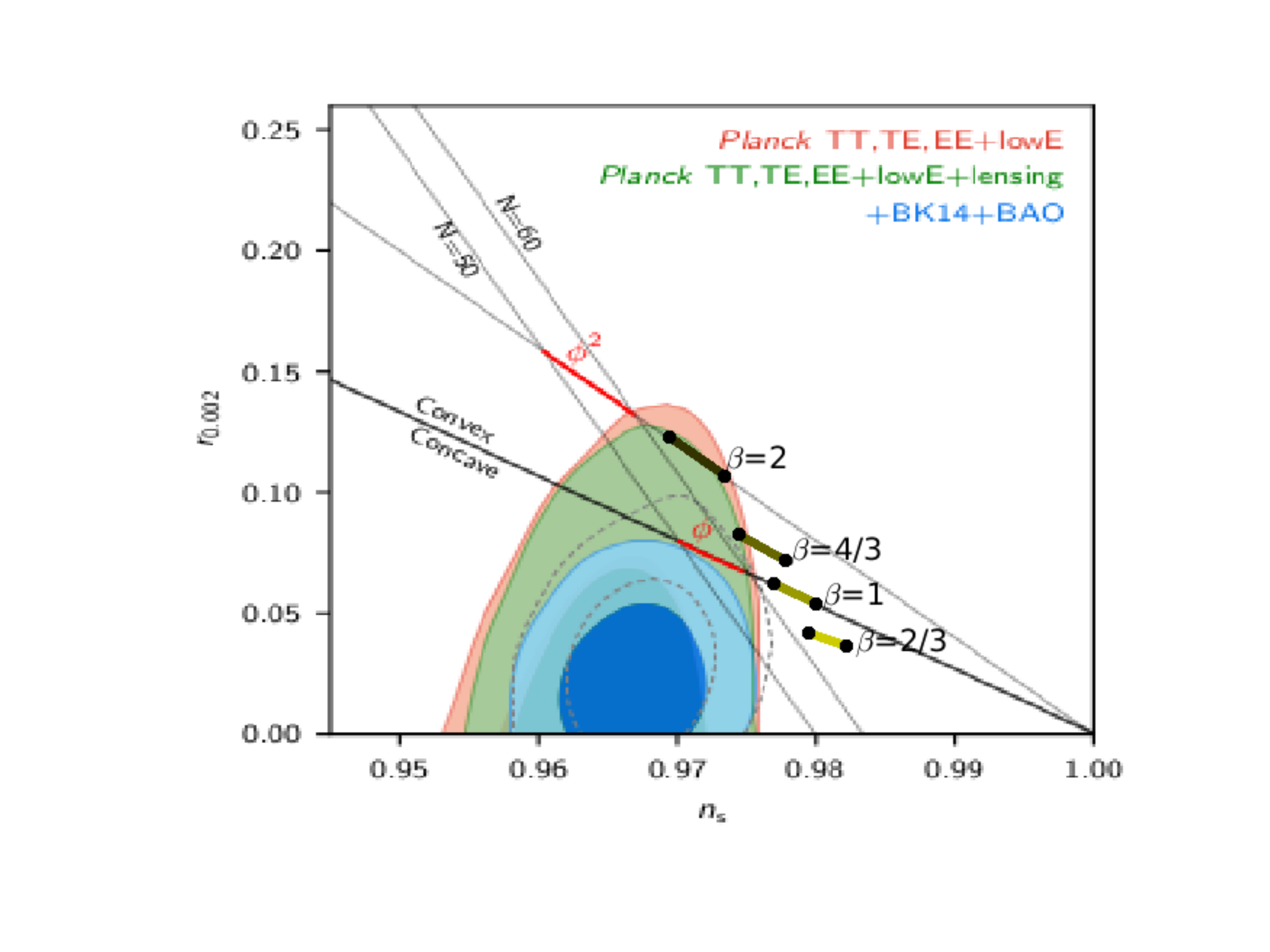}
%\end{center}
\caption{Marginalized joint confidence contours for $(n_s,r)$ at $68\%$ and $95\%$ confidence level.  Considering the inflationary piece of the potential as 
$V=\lambda \phi^{\beta}$ \cite{rp}, in
quintessential inflation, for the values of $\beta=2, 3/4, 1, 2/3$, we have
drawn the curves  from  $65$ to $75$ e-folds (see the black curves). And  when one considers the standard inflation, for $\beta=2, 1$, the curves have been drawn in red from $50$ to $60$ $e$-folds.
As one can see that the quadratic potential ($V \propto \phi^2$), which is disregarded in standard inflation is not disfavored in quintessential inflation
(Figure courtesy of the Planck 2018 collaboration  \cite{planck18a}).}
\label{fig:n-r}
\end{figure}

To end this Section some remarks are in order:

{\it Remark 1:} One could also choose a very general PV-type potentials:
\begin{eqnarray}
V_{\alpha}(\varphi)=\left\{\begin{array}{ccc}
\lambda(\varphi^{\beta}-M_{pl}^{\beta} + M^{\beta}) & \mbox{for} & \varphi\leq -M_{pl}\\
\lambda\frac{M^{\alpha+\beta}}{(\varphi+M_{pl})^{\alpha}+M^{\alpha}} &\mbox{for} & \varphi\geq -M_{pl},\end{array}
\right.
\end{eqnarray}
with $1<\beta\leq 2$ 
to fit  with  Planck 2018 observational data \cite{planck18}. 
As one can see from Fig. \ref{fig:n-r} that if one considers the number of $e$-folds from the moment that the pivot scale  crosses the Hubble radius to the end of inflation, is, for example,  between $65$ and $75$, which  usually happens in quintessential inflation due to the kination period after inflation \cite{deHaro:2017nui},  then the improved PV-model (\ref{improved})  enters in the $2\sigma$ C.L. for the  Planck likelihoods: Planck TT, TE, EE + low E, and 
Planck TT, TE, EE + low E + lensing.  
On the other hand, if one  takes into account the ``{\it tensor sector}'' and  wants that a quintessential model matches with  the   Planck TT, TE, EE + low E+ lensing + BK14 + BAO likelihood, one has to consider other kind of 
potentials, such as  plateau potentials \cite{Geng:2017mic} or  $\alpha$-attractors \cite{Dimopoulos,  akrami}.

{\it Remark 2:} Potentials with a sudden break have interesting properties. For example, in \cite{starobinsky92} a model with a local singularity of the potential in the slow-roll phase,
was considered  to  study the deviation  from a flat power spectrum. For our 
 potential (\ref{improved}), the first derivative has a discontinuity at $\varphi=-M_{pl}$   which enhances the gravitational production of superheavy  particles as was shown in \cite{Chung}. In fact, the greatest is the order of the first discontinuous derivative the lower is the energy density of the superheavy gravitationally produced particles, and thus the lower is the reheating temperature.  This means  that, in quintessential inflation,  for smooth potentials such as plateau potentials or $\alpha$-attractors  the particle creation has to be done via instant preheating \cite{fkl}.  On the other hand, as has been recently suggested, the abundance of dark matter could be explained via the production of
only gravitationally interacting massive particles (GIMP) \cite{Hashiba1,Chung1,Ema,Chung2}, which could not be applied to smooth potentials, however,  this mechanism of production of dark matter seems to work very well in PV models \cite{haro19}.

{\it Remark 3:} What would be interesting, although is a point that deserves future investigation,  is to find models whose inflationary potential  was a plateau-type potential to enter in the Planck TT, TE, EE + low E+ lensing + BK14 + BAO likelihood, matching abruptly  (not smoothly) with an quintessence potential in order that the superheavy dark matter
was created gravitationally. For example, using an Exponential  SUSY inflation-type potential
\begin{eqnarray}
V_{\alpha}(\varphi)=\left\{\begin{array}{ccc}
\lambda M_{pl}^4\left( 1-e^{\alpha\varphi/M_{pl}} + \left(\frac{M}{M_{pl}}\right)^4\right) & \mbox{for} & \varphi\leq 0\\
\lambda\frac{M^8}{\varphi^{4}+M^4} &\mbox{for} & \varphi\geq 0,\end{array}
\right.
\end{eqnarray}
or, a Higgs Inflation-type potential
\begin{eqnarray}
V_{\alpha}(\varphi)=\left\{\begin{array}{ccc}
\lambda M_{pl}^4\left( 1-e^{\alpha\varphi/M_{pl}} + \left(\frac{M}{M_{pl}}\right)^2\right)^2 & \mbox{for} & \varphi\leq 0\\
\lambda\frac{M^8}{\varphi^{4}+M^4} &\mbox{for} & \varphi\geq 0.\end{array}
\right.
\end{eqnarray}

For both potentials one can calculate that,
\begin{eqnarray}
n_s\cong 1-\frac{2}{N},\qquad r\cong \frac{8}{\alpha^2 N^2},
\end{eqnarray}
which directs that for $\alpha$ of the order of $1$ and for a number of $e$-folds greater than $60$, the ratio of tensor to scalar perturbations is less than $0.003$. Thus, the  spectral index and the tensor/scalar ratio 
enter perfectly in the  two dimensional marginalized joint confidence contour at $1\sigma$ CL for the Planck TT, TE, EE + low E+ lensing + BK14 + BAO likelihood.

\section{Evolution from kination to matter-radiation equality}
\label{sec-kin-to-mr}

During the phase transition from inflation to kination the particles are produced. When the produced particles have very heavy masses, which occur in instant preheating or when one only considers heavy massive particles conformally coupled to gravity, these particles have to decay into light particles to form a relativistic plasma in thermal equilibrium whose energy density will eventually be dominant in order to match with the hot Friedmann universe. Two different cases may arise: when the decay of the heavy massive particles occurs before or after the end of the kination regime. Here, we will study both cases separately. It is also important to take into account that during this period the  potential energy of the inflaton field, due its low value compared with the kinetic energy could be safely disregarded, which implies that the dynamical equations could be solved analytically.

\subsection{Decay  before the end of kination}
\label{subsec-before-the-end-kin}

Assuming that the produced massive particles at the phase transition decay into lighter ones before the equality between the energy density of the inflaton and those of the created particles, at the reheating time, one has 
\begin{eqnarray}
\varphi_{rh}=\varphi_{kin}+\sqrt{\frac{2}{3}}M_{pl}\ln\left( \frac{H_{kin}}{H_{rh}} \right),
\end{eqnarray}
and using that at the reheating time, i.e., when the energy density of the scalar field and the one of the relativistic plasma coincide: $H_{rh}^2=\frac{2\rho_{rh}}{3M_{pl}^2}$, one gets 
\begin{eqnarray}
\varphi_{rh}=\varphi_{kin}+\sqrt{\frac{2}{3}}M_{pl}\ln\left( \frac{\sqrt{\frac{1+\sqrt{3}}{6}}m}{\sqrt{\frac{\pi^2g_{rh}}{45}} \frac{T_{rh}^2}{M_{pl}}}\right),
\qquad \mbox{and}  \qquad \dot{\varphi}_{rh}=\sqrt{\frac{6\pi^2g_{rh}}{45}} T_{rh}^2,\end{eqnarray} 
where we have used that at reheating time, the energy density and the temperature are related via 
 $\rho_{rh}=\frac{\pi^2}{30}g_{rh}T_{rh}^4$, where the number of degrees of freedom is, $g_{rh}=107$  \cite{rg}.

 We consider reheating via  gravitational particle production of massive particles 
obtaining a reheating temperature is of the
 order $T_{rh}=100$ TeV \cite{haro18}
and via instant preheating leading to a reheating temperature  of the order $T_{rh}=10^9$ GeV \cite{fkl, fkl0}.

In the former case one has  \begin{eqnarray}\varphi_{rh}\cong 37.8 M_{pl}, \qquad \dot{\varphi}_{rh}\cong 2\times 10^{-26} M_{pl}^2,
 \end{eqnarray}
 and  when  reheating via instant preheating is considered one gets
\begin{eqnarray}\varphi_{rh}\cong 22.8 M_{pl}, \qquad \dot{\varphi}_{rh}\cong 2\times 10^{-18} M_{pl}^2.
 \end{eqnarray} 
During the radiation dominated phase, one can continue disregarding the gradient of the potential,  obtaining
 \begin{eqnarray}
 \varphi(t)=\varphi_{rh}+2\dot{\varphi}_{rh}t_{rh}\left( 1-\sqrt{\frac{t_{rh}}{t}} \right),
 \end{eqnarray}
 and we have to calculate the value of the field and its derivative at the matter-radiation equality, i.e., $\varphi_{eq}$ and $\dot{\varphi}_{eq}$ which will be the initial conditions of our dynamical system.

To do it, we consider the central values obtained in \cite{planck} (see the second column in Table $4$) of  the red shift at the matter-radiation equality $z_{eq}=3365$,
the present value of the ratio of the matter energy density to the critical one $\Omega_{m,0}=0.308$, and $H_0=67.81\; \mbox{Km/sec/Mpc}=5.94\times 10^{-61} M_{pl}$.
Then, the present value of the matter energy density is $\rho_{m,0}=3H_0^2M_{pl}^2\Omega_{m,0}=3.26\times 10^{-121} M_{pl}^4$, and at matter-radiation equality we will 
have $\rho_{eq}=2\rho_{m,0}(1+z_{eq})^3=2.48\times 10^{-110} M_{pl}^4=8.8\times 10^{-1} \mbox{eV}^4$. 
Now,  using the relation at the matter-radiation equality $\rho_{eq}=\frac{\pi^2}{15}g_{eq}T_{eq}^4$ with $g_{eq}=3.36$ (see \cite{rg}), we  get 
$T_{eq}=3.25\times 10^{-28} M_{pl}=7.81\times 10^{-10}$ GeV.

\

\begin{remark}
It is not mandatory to use the observational values of the parameters ($z_{eq}$, $\Omega_{m,0}, H_0)$, since one can use  the observational values of other three parameters. For example, using $H_0$, $\Omega_{m,0}$, and $T_0$,  one has $\rho_{m,0}=3H_0^2M_{pl}^2\Omega_{m,0}$ and since $\rho_{m,0}=\frac{\rho_{eq}}{2}\left( \frac{a_{eq}}{a_0}\right)^3$
and $\rho_{eq}=\frac{\pi^2}{15}g_{eq} T^4_{eq}=\frac{\pi^2}{15}g_{eq} T^4_{0}\left( \frac{a_{0}}{a_{eq}}\right)^4$ one gets
\begin{eqnarray}
\rho_{m,0}=\frac{\pi^2}{30}g_{eq} T^4_{0}\left( \frac{a_{0}}{a_{eq}}\right)=\frac{\pi^2}{30}g_{eq} T^4_{0}\left( z_{eq}+1\right) \Longrightarrow
z_{eq}=-1+\frac{90}{g_{eq}\pi^2}\frac{H_0^2M_{pl}^2}{T_0^4}\Omega_{m,0}.
\end{eqnarray}
Thus, the observational value of $z_{eq}$ is obtained from the observational values of $H_0$, $\Omega_{m,0}$  and the well-known current temperature of the universe
$T_0$.
\end{remark}

In this way the observational values of $H_0$ and $\Omega_{m,0}$  could be obtained directly from the own  $V_{\alpha}$ potential, however, the parameter $M$ is completely degenerate as reported from the latest astronomical datasets \cite{haro18}, where the addition of baryon acoustic oscillations data into the cosmic microwave background radiation cannot break such a degeneracy. The low redshifts sample like Pantheon  from the Supernovae Type Ia, and the Hubble parameter measurements
from the cosmic chronometers also return similar conclusion. 
For $V_4$ (that means, $V_{\alpha}$ when $\alpha =4$) the central values of these parameters are (see the third column of the table $3$ of \cite{haro18}) $\Omega_{m,0}=0.306$
and $H_0=67.92 \frac{km/sec}{Mpc}\cong 5.95 \times 10^{-61} M_{pl}$. Then, taking into account that $T_0\cong 2.7$ K  $\cong 2.33\times 10^{-13}$ GeV
$\cong 9.7\times 10^{-32} M_{pl}$ one gets, 
$z_{eq}=3321$.

 \begin{enumerate}
 \item Reheating via production of heavy  massive particles conformally coupled to gravity:
 
\begin{align}\label{eq}
 \varphi_{eq} =\varphi_{rh}+2\sqrt{\frac{2}{3}}M_{pl}\left(1-\sqrt{\frac{2H_{eq}}{3H_{rh}}}\right)
 =\varphi_{rh}+2\sqrt{\frac{2}{3}}M_{pl}\left(1-\sqrt{\frac{2}{3}}\left( \frac{g_{eq}}{g_{rh}} \right)^{1/4}\frac{T_{eq}}{T_{rh}}\right) \nonumber \\
\cong \varphi_{rh}+2\sqrt{\frac{2}{3}}M_{pl} 
  \cong 39.4 M_{pl}.
  \end{align}

 \begin{align}\label{doteq}
\dot{\varphi}_{eq}=\dot{\varphi}_{rh}\frac{t_{rh}}{t_{eq}}\sqrt{\frac{t_{rh}}{t_{eq}}}=\frac{4}{3}M_{pl}H_{eq}\sqrt{\frac{H_{eq}}{H_{rh}}} 
=\frac{4\pi}{9}\sqrt{\frac{g_{eq}}{5}}\left(\frac{g_{eq}}{g_{rh}} \right)^{1/4}\frac{T_{eq}^3}{T_{rh}}\cong  2.3\times 10^{-15} \mbox{ eV}^2,
\end{align} 

\item Instant preheating:

\begin{eqnarray}
 \varphi_{eq}=\varphi_{rh}+2\sqrt{\frac{2}{3}}M_{pl}\left(1-\sqrt{\frac{2H_{eq}}{3H_{rh}}}\right)
 =\varphi_{rh}+2\sqrt{\frac{2}{3}}M_{pl}\left(1-\sqrt{\frac{2}{3}}\left( \frac{g_{eq}}{g_{rh}} \right)^{1/4}\frac{T_{eq}}{T_{rh}}\right) \nonumber \\
  \cong \varphi_{rh}+2\sqrt{\frac{2}{3}}M_{pl} 
  \cong 24.4 M_{pl}.
  \end{eqnarray}

 \begin{align}
\dot{\varphi}_{eq}=\dot{\varphi}_{rh}\frac{t_{rh}}{t_{eq}}\sqrt{\frac{t_{rh}}{t_{eq}}}=\frac{4}{3}M_{pl}H_{eq}\sqrt{\frac{H_{eq}}{H_{rh}}}
=\frac{4\pi}{9}\sqrt{\frac{g_{eq}}{5}}\left(\frac{g_{eq}}{g_{rh}} \right)^{1/4}\frac{T_{eq}^3}{T_{rh}}\cong  2.3\times 10^{-19} \mbox{ eV}^2.
\end{align}

\end{enumerate}

\begin{remark} During radiation dominated era, the potential energy of the field is negligible, so since
$\dot{\varphi}=\dot{\varphi}_{rh}\left( \frac{t_{rh}}{t} \right)^{3/2}=\dot{\varphi}_{rh}\left( \frac{2H(t)}{3H_{rh}} \right)^{3/2}$, we will have
$\rho_{\varphi}(t)=\frac{\dot{\varphi}^2}{2}=\rho_{rh}\left( \frac{2H(t)}{3H_{rh}} \right)^{3}$. On the other hand, the energy density of the background is $\rho(t)=\rho_{rh}\left( \frac{2H(t)}{3H_{rh}} \right)^{2}$,
then the ratio of the energy density of the scalar field to the energy density of the background is

\begin{eqnarray}\Omega_{\varphi}(t)=\frac{\rho_{\varphi}(t)}{\rho_c(t)}=\frac{2 H(t)}{3H_{rh}}= \frac{\sqrt{2} T^2(t)}{3T^2_{rh}}. 
\end{eqnarray}

\end{remark}
Since the Big Bang Nucleosynthesis  occurs at temperatures about $1$ MeV, for a reheating temperature, $T_{rh}=100$ TeV,  we have $\Omega_{\varphi}(t_{BBN})\sim 10^{-16}$. However,  if one wants a greater value of this parameter one has to consider lower reheating temperatures  satisfying  $T_{rh}\gtrsim1$ MeV.

\subsection{Decay  after the end of kination}

In this case, which is only possible when reheating is due to the production of heavy massive particles, (in the case of instant preheating this will could
produce an undesired  new inflationary era \cite{fkl, haro18a}),
 it is possible to obtain reheating temperature very close to $1$ MeV \cite{haro18}. So, here we consider $T_{rh}=1$ MeV, which means that 
 at the BBN epoch,  $\Omega_{\varphi}(t_{BBN})=\frac{{2}}{3}\cong 0.66$, and at the matter-radiation equality, $\Omega_{\varphi, eq}\sim 10^{-13}$.

Let $\bar{t}$ be the time at which kination ends,  that is, when $\rho_{\varphi}(\bar{t})=\rho_{\chi}(\bar{t})$, where $\rho_{\chi}$ denotes the energy density of the produced massive 
$\chi$-particles. Introducing the {\it heat efficiency} defined by  $\Theta=\frac{\rho_{\chi, kin}}{\rho_{\varphi, kin}}$  \cite{rubio}, and taking into account that
\begin{eqnarray}
\rho_{\varphi}(\bar{t})=\rho_{\varphi,kin}\left(\frac{a_{kin}}{a(\bar{t})}\right)^6, \quad \rho_{\chi}(\bar{t})=\rho_{\chi, kin}\left(\frac{a_{kin}}{a(\bar{t})}\right)^3~, \quad \mbox{and} \quad   H(\bar{t})=\frac{2\rho_{\varphi}(\bar{t})}{3M_{pl}^2},
\end{eqnarray} 
one may deduce that $H(\bar{t})=\sqrt{2}H_{kin}\Theta$. 
Now since  during kination, $H({t})=\frac{1}{3{t}}$, one can conclude that 
$\bar{t}=\frac{1}{3\sqrt{2}H_{kin}\Theta}=\frac{1}{\sqrt{3(1+\sqrt{3})}m\Theta}$.

On the other hand, at the end of kination we will have
\begin{eqnarray}
\varphi(\bar{t})=\varphi_{kin}+\sqrt{\frac{2}{3}}M_{pl}\ln\left(\frac{\bar{t}}{t_{kin}} \right)=\varphi_{kin}-\sqrt{\frac{2}{3}}M_{pl}\ln\left(\sqrt{2}\Theta \right),
\end{eqnarray}
and
\begin{eqnarray}
\dot{\varphi}(\bar{t})=\sqrt{\frac{2}{3}}\frac{M_{pl}}{\bar{t}}=\sqrt{2(1+\sqrt{3})}mM_{pl}\Theta.
\end{eqnarray}

During the period between $\bar{t}$ and $t_{rh}$, the universe is matter dominated, and thus the Hubble parameter becomes, $H=\frac{2}{3t}$. Since the gradient of the potential could be disregarded at this
epoch,  hence, the equation of the scalar field becomes, $\ddot{\varphi}+\frac{2}{t}\dot{\varphi}=0$, and thus, at the reheating time
\begin{eqnarray}
\varphi_{rh}=\varphi(\bar{t})+\sqrt{\frac{2}{3}}M_{pl}\left( 1-\frac{\bar{t}}{t_{rh}}  \right)=
\varphi(\bar{t})+\sqrt{\frac{2}{3}}M_{pl}\left( 1-\frac{H_{rh}}{2\sqrt{2}H_{kin}\Theta}\right)\nonumber\\ =
\varphi(\bar{t})+\sqrt{\frac{2}{3}}M_{pl}\left( 1-\frac{\frac{\pi}{6}\sqrt{\frac{g_{rh}}{10}}\frac{T_{rh}^2}{M_{pl}}}{\sqrt{\frac{1+\sqrt{3}}{\sqrt{3}}}m\Theta}\right)\end{eqnarray}

\begin{eqnarray}
\dot{\varphi}_{rh}=\sqrt{\frac{2}{3}}\frac{M_{pl}\bar{t}}{t_{rh}^2}=\frac{\sqrt{3}}{4}\frac{M_{pl}H_{rh}^2}{H_{kin}\Theta}
=\frac{\sqrt{2}\pi^2}{120\sqrt{1+\sqrt{3}}}\frac{g_{rh}T_{rh}^4}{mM_{pl}\Theta}.\end{eqnarray}

\

\textit{Calculation of the heat efficiency $\Theta$:} The energy density of the produced massive particles with mass $m_{\chi}=5\times 10^{-4}M_{pl}$ is given by
$\rho_{\chi,kin}\cong 10^{-5}\left(\frac{m}{m_{\chi}} \right)^2m^4= 6.25\times 10^{-31} M_{pl}^4$ \cite{haro18} and the one of the inflaton field is
$\rho_{\varphi, kin}=3H_{kin}^2M_{pl}^2= \frac{1+\sqrt{3}}{2}m^2M_{pl}^2\cong 3.41\times 10^{-11} M_{pl}^4$, thus $\Theta=1.83\times 10^{-20}$, which means that
\begin{eqnarray}
\varphi(\bar{t})\cong 37.37 M_{pl}\quad \mbox{and} \quad  \dot{\varphi}(\bar{t})\cong 7.07 \times 10^{-26} M_{pl}^2.
\end{eqnarray}

For a reheating temperature of the order of  $1$ MeV   we will have $\left(\frac{T_{rh}^2}{M_{pl}}\right)/m\Theta \sim 10^{-16}$, and thus,
\begin{eqnarray}
\varphi_{rh}\cong \varphi(\bar{t})+\sqrt{\frac{2}{3}}M_{pl}\cong 38.18 M_{pl}\quad \mbox{and} \quad \dot{\varphi}_{rh}\cong 0.
\end{eqnarray}

Finally, using the relations in eqn. (\ref{eq})  one gets,
 $\varphi_{eq}\cong 39.81$ and $\dot{\varphi}_{eq}\cong 0$. In the next section, we shall describe the evolution of the universe after the matter-radiation equality to present.

\section{Evolution from the matter-radiation equality}
\label{sec-mat-rad-equality}

After the matter-radiation equality, the dynamical equations could not be solved analytically and thus, one needs to use numerics, starting at the
matter-radiation equality, to compute them. In order to do that we need to use a ``time'' variable  that we choose to be the number of $e$-folds up to the present epoch, namely, $N\equiv -\ln(1+z)=\ln\left( \frac{a}{a_0}\right)$. Now, using  the variable $N$,  one can recast the  energy density of radiation and matter respectively as, 
\begin{eqnarray}
\rho_{r}(a)=\frac{\rho_{eq}}{2}\left(\frac{a_{eq}}{a}  \right)^4\Longrightarrow \rho_{r}(N)= \frac{\rho_{eq}}{2}e^{4(N_{eq}-N)} ,
\end{eqnarray}
and
\begin{eqnarray}
\rho_{m}(a)=\frac{\rho_{eq}}{2}\left(\frac{a_{eq}}{a}  \right)^3\Longrightarrow \rho_{m}(N)=\frac{\rho_{eq}}{2}e^{3(N_{eq}-N)},
\end{eqnarray}
where  the value of the energy density at the matter-radiation equality { $\rho_{eq}\cong 8.8\times 10^{-1}$  ${\mbox eV}^4$}, 
has been obtained in the previous Section \ref{sec-kin-to-mr} (precisely see the subsection \ref{subsec-before-the-end-kin}) and also one can understand that $N_{eq}$ is the value of $N$ at the matter-radiation equality.

Now, in order to obtain the dynamical system for this scalar field model, we 
introduce the following dimensionless variables
 \begin{eqnarray}
 x=\frac{\varphi}{M_{pl}}, \qquad y=\frac{\dot{\varphi}}{K M_{pl}},
 \end{eqnarray}
 where $K$ is a parameter (with some dimension) that we will determine right now. Now, using the variable, 
 $N = - \ln (1+z)$, defined above and also using the conservation equation $\ddot{\varphi}+3H\dot{\varphi}+V_{\varphi}=0$, one can construct the following  non-autonomous dynamical system:
 \begin{eqnarray}\label{system}
 \left\{ \begin{array}{ccc}
 x^\prime & =& \frac{y}{\bar H}~,\\
 y^\prime &=& -3y-\frac{\bar{V}_x}{ \bar{H}}~,\end{array}\right.
 \end{eqnarray}
 where the prime represents the derivative with respect to $N$, $\bar{H}=\frac{H}{K}$   and $\bar{V}=\frac{V}{K^2M_{pl}^2}$. Moreover, the Hubble equation now looks as  
 \begin{eqnarray}
 \bar{H}=\frac{1}{\sqrt{3}}\sqrt{ \frac{y^2}{2}+\bar{V}(x)+ \bar{\rho}_{r}(N)+\bar{\rho}_{m}(N) }~,
 \end{eqnarray}
where we have introduced the following dimensionless energy densities
 $\bar{\rho}_{r}=\frac{\rho_{r}}{K^2M_{pl}^2}$ and 
 $\bar{\rho}_{m}=\frac{\rho_{m}}{K^2M_{pl}^2}$.

We choose  $K\cong 4.1\times 10^{-32}$ eV, in order to have $KM_{pl}\cong 10^{-4} \mbox{eV}^2$, and
we take the following initial conditions at matter-radiation equality: 
\begin{enumerate}
\item For the reheating temperature $T_{rh}=100$ TeV  (the heavy massive particles decay before the end of kination):

$x_{eq}=39.4$ and $y_{eq}=2.3\times 10^{-11}$.

\item For the reheating temperature $T_{rh}=10^9$ GeV (instant preheating):

$x_{eq}=24.4$ and $y_{eq}=2.3\times 10^{-15}$.

\item For the reheating temperature $T_{rh}=1$ MeV (the heavy massive particles decay after the end of kination):

 $x_{eq}=39.81$ and $y_{eq}=0$.
 
\end{enumerate}

For a quick look at the initial conditions at different reheating temperatures, in Table \ref{tab:ic} we have summarized them.  

\begin{table*}                                                                                                                
\begin{tabular}{ccccc}                                                                                                            
\hline\hline                                                                                                                    
Serial No. &~~  Reheating Temperature ($T_{rh}$) & $(x_{eq}, y_{eq})$\\ \hline
1 & $100$ TeV & $(39.4, 2.3\times 10^{-11})$ \\
2 & $10^9$ GeV & $(24.4, 2.3\times 10^{-15})$ \\
3 & $1$ MeV   & $(39.81, 0)$ \\

\hline\hline                                                                                                                    
\end{tabular}                                                                                                                   
\caption{Summary of the initial conditions at the matter-radiation equality at different reheating temperatures is shown. }\label{tab:ic}                                                                                                   
\end{table*}

On the other hand, 
\begin{eqnarray}
\bar{\rho}_{r}(N)= 4.4 \times 10^{7} e^{4(N_{eq}-N)}, \qquad \bar{\rho}_{m}(N)= 4.4\times  10^{7}  e^{3(N_{eq}-N)},
\end{eqnarray}
with  $N_{eq}\cong -8.121$ and  $\bar{H}_0\cong 3.53\times 10^{-2}$ .

To integrate the dynamical system for the potential $V_4$, writing $M=\bar{M} \times10^{-18} M_{pl}=2.4 \bar{M}$ GeV, one gets,

\begin{eqnarray}
\bar{V}_4(x)=\frac{25}{2}\frac{(2.4)^4 \bar{M}^6\times 10^{-4}}{(x+1)^4+ \bar{M}^4\times10^{-72}}\cong 4.1\times 10^{-2} \frac{\bar{M}^6}{(x+1)^4}.
\end{eqnarray}

Then, to find the accurate value of $M$, which  depends on the reheating temperature, we have solved numerically the dynamical system (\ref{system}) with initial conditions $x_{eq}$ and $y_{eq}$ at $N_{eq}=-8.121$, for different values of $\bar{M}$, which leads to the value $\bar{H}_0=3.53\times 10^{-2}$.  Numerically, we have obtained that, 

\begin{enumerate}
 \item $T_{rh}=100$ TeV,   $\bar{M}=7.43 \Longrightarrow M=18.1$ GeV.  
  \item $T_{rh}=10^9$ GeV,  $\bar{M}=5.46 \Longrightarrow M=13.3$ GeV.  
  \item $T_{rh}=1$ MeV,  $\bar{M}=7.48\Longrightarrow M=18.3$ GeV.  
  \end{enumerate}
  
\begin{enumerate}
     
\item For the potential $V_2$  taking $M=\bar{M}$ eV, one gets
$ \bar{V}_2(x)\cong 1.2\times 10^{-3} \frac{\bar{M}^4}{(1+x)^2}$. Numerically, we have obtained that, $\bar{M}=7.70,\; 6.11,\; 7.74$, for  $T_{rh} = 100$ TeV, $T_{rh} = 10^9$ GeV, and  $T_{rh } = 1$ MeV respectively.

\item In the same way,  taking $M=10^{2} \bar{M}$ TeV, one has
$ \bar{V}_6(x)\cong 3.7\times 10^{-1} \frac{\bar{M}^8}{(1+x)^6}$. And by numerical calculations, we obtained that, $\bar{M}=8.62,\; 6.09,\; 8.69$, for  $T_{rh} = 100$ TeV, $T_{rh} = 10^9$ GeV, and  $T_{rh } = 1$ MeV respectively.

\end{enumerate}

We now focus on the qualitative evolution of the cosmological parameters in terms of their graphical behavior by solving the corresponding equations numerically. In order to do so, we further introduce another cosmological parameter which is the   
the effective Equation of State (EoS) parameter, given by 
\begin{eqnarray}
w_{eff} \equiv -1-\frac{2\dot{H}}{3H^2}=-1-\frac{2{H}'}{3H}= -1-\frac{2\bar{H}'}{3\bar{H}},
\end{eqnarray}
where we have used that $\dot{H}=H'H$. Here, the `overdot' and prime, as already mentioned earlier, represent the differentiation with respect to the cosmic time and $N =\ln (a/a_0)$, respectively. 

Now, in order to depict the evolution of the cosmological parameters, we have 
only considered the case $\alpha=4$ and reheating temperature $T_{rh}= 100$ TeV because the other $\alpha$-cases and reheating temperatures lead to practically the same results. Let us define that the density parameter for the scalar field is, $\Omega_{\varphi}(N)=\frac{\rho_{\varphi}}{3H^2M_{pl}^2}=
\frac{\bar{\rho}_{\varphi}}{3\bar{H}^2}$, where $\bar{\rho}_{\varphi}=y^2/2+\bar{V}_{\alpha}(x)$, while the density parameters for the matter sector and the radiation respectively take the expressions,  
$\Omega_{m}(N)=\frac{\bar{\rho}_{m}}{3\bar{H}^2}$ and $\Omega_{r}(N)=\frac{\bar{\rho}_{r}}{3\bar{H}^2}$.

\begin{figure}
\begin{center}
\includegraphics[scale=0.45]{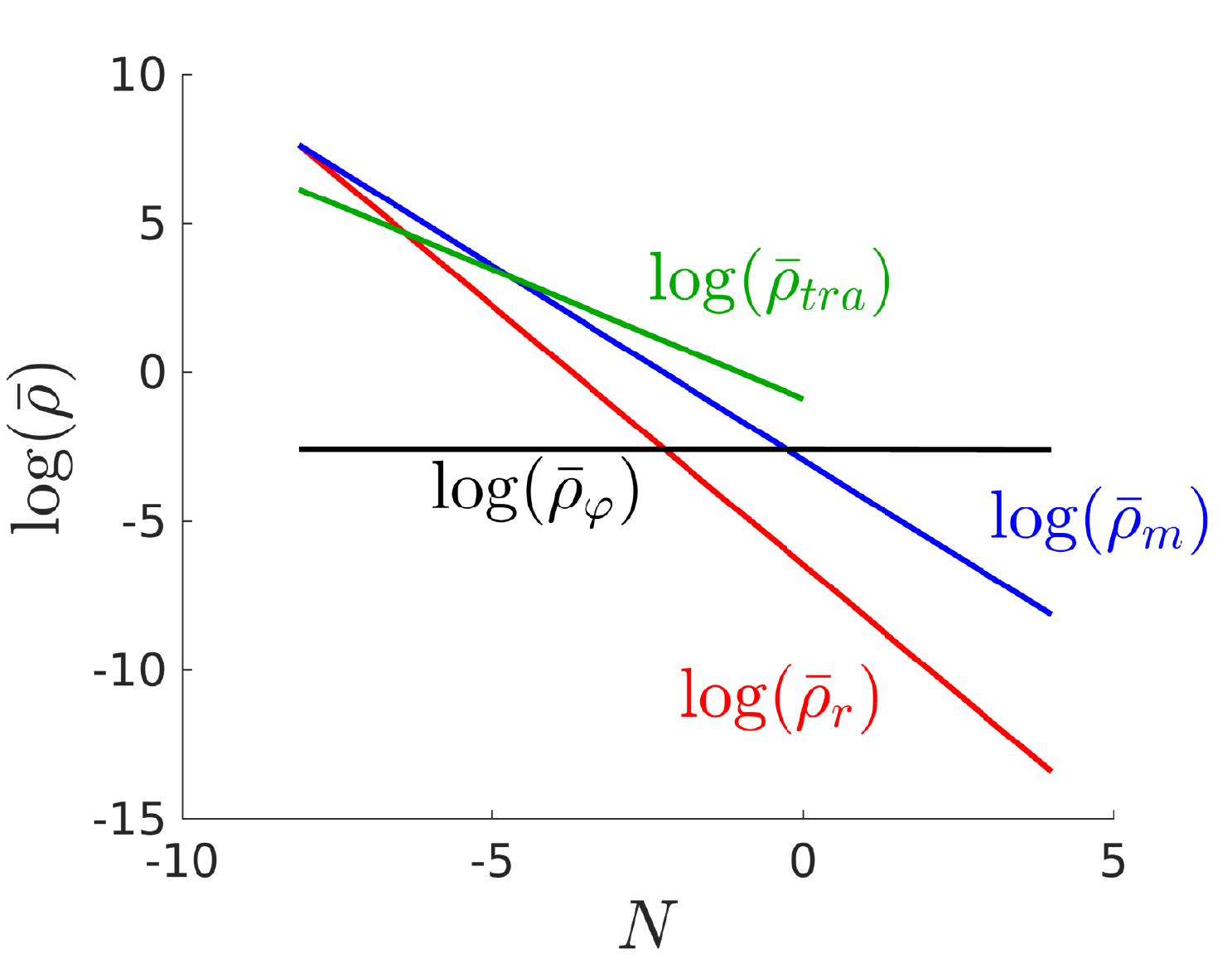}
\end{center}
\caption{(Color online) Qualitative evolutions of different energy densities have been presented for $\alpha=4$, rehating temperature $T_{rh} = 100$ TeV and $M=18.1$ GeV. The blue, red, green and  black curves respectively present the matter, radiation, tracker solution and the  scalar field where all of them are considered in the logarithmic units. The initial conditions for all the curves are taken at the matter-radiation equality.}
\label{fig:modifiedPV1}
\end{figure}
\begin{figure}
\begin{center}
\includegraphics[scale=0.45]{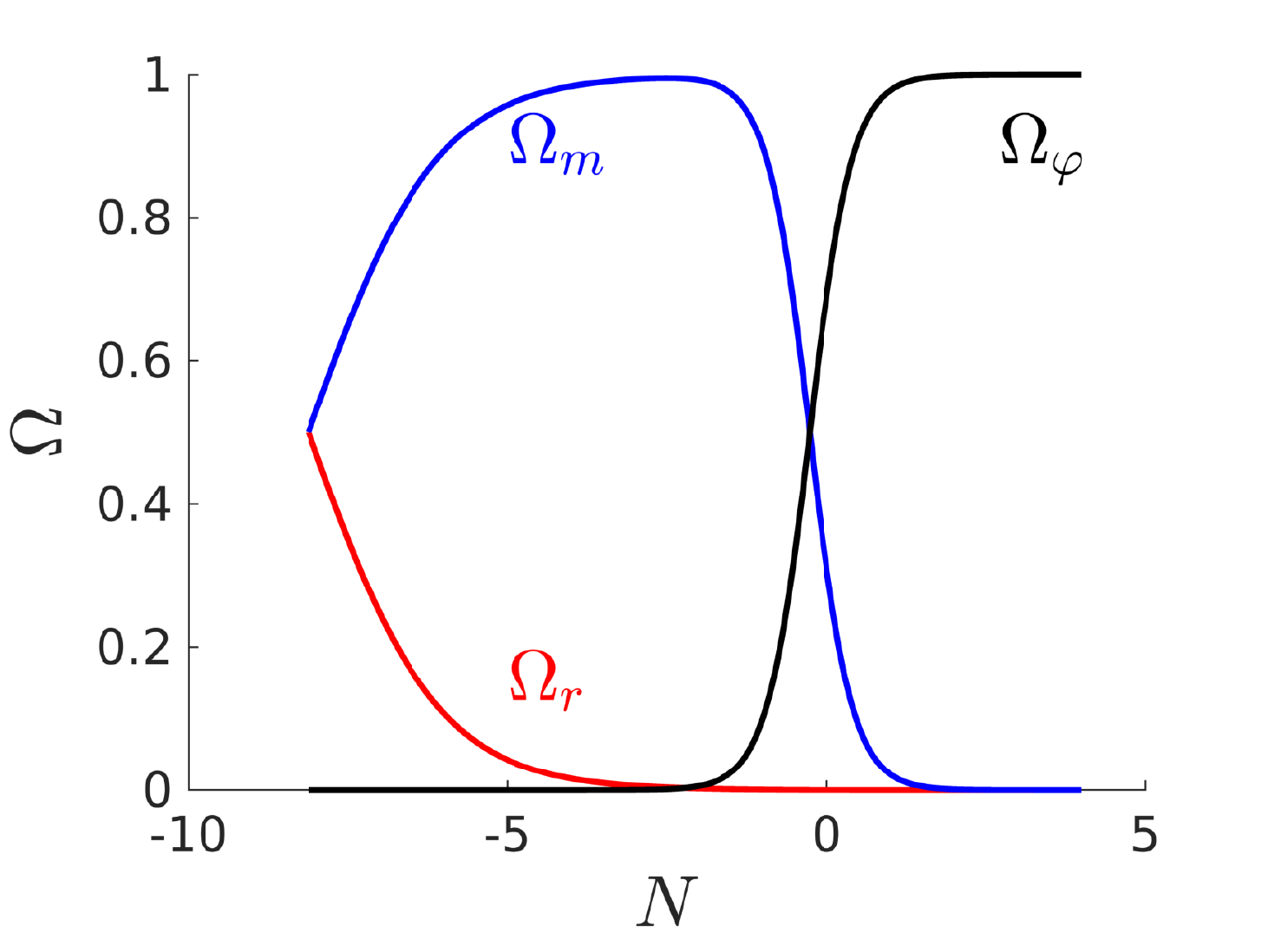}
\includegraphics[scale=0.45]{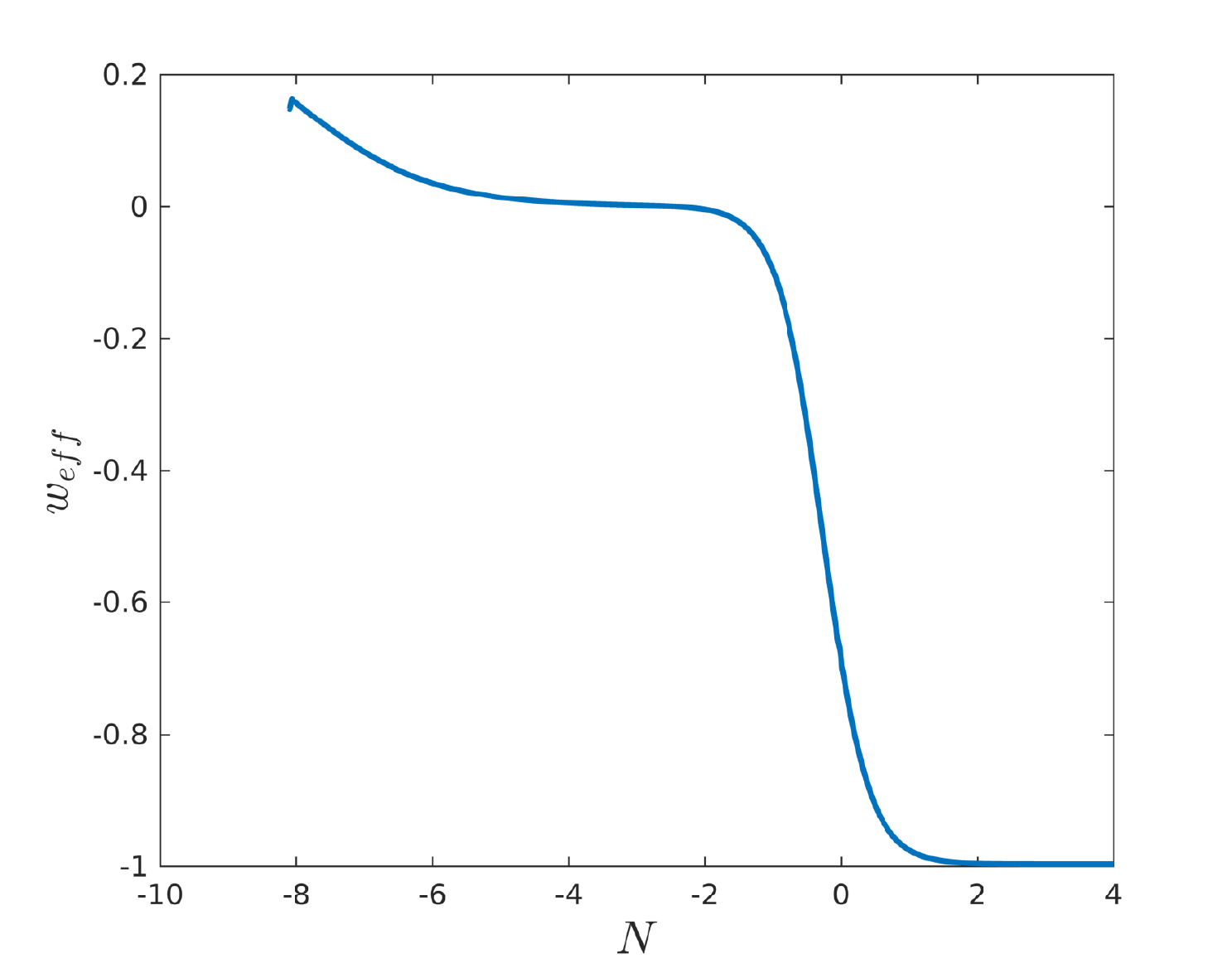}
\end{center}
\caption{(Color online) In the left panel we show the evolution of the density parameter $\Omega$ for matter (blue curve), radiation (red curve), and the scalar field (black curve) while in the right panel we show the evolution of the effective equation of state parameter, $w_{eff}$ (blue curve). In all the cases, during the numerical simulations, we assume $\alpha=4$, reheating temperature $T_{rh} =100$ TeV and $M=18.1$ GeV. From both the left and right graphs, one can roughly estimate that at the present time $\Omega_{\varphi} \simeq 0.7$, $\Omega_{m} \simeq 0.3$, $\Omega_{r} \simeq 0.0$ and $w_{eff} \simeq -0.8$, while at late time, the respective quantities 
tend to $1,0,0$ and $-1$. Moreover, from the graph of $w_{eff}$ (right panel) one can see that after the matter-radiation equality the universe enters into the matter dominated era (i.e., $w_{eff}=0$) and then subsequently enters into the present accelerated regime and asymptotically approaches toward $w_{eff}  = -1$. }
\label{fig:modifiedPV2}
\end{figure}

On the other hand, one can also recall 
the relation between the cosmic time $t$ and the time $N$ as follows 
\begin{eqnarray}
t(N)-t_0=\frac{1}{K}\int_0^N\frac{d\tau}{\bar{H}(\tau)},
\end{eqnarray}
where $t_0$ denotes the present cosmic time. Now, using the above relation, one can calculate many important things in the following way. 

From the numerical simulations, for the fixed values of $\alpha$ ($= 4$) and reheating temperature $T_{rh}=100$ TeV, we find that the epoch describing the phase $\rho_m=\rho_{\varphi}$ happened at
$N= -0.28$ which in terms of the cosmic time gives, $ t-t_0= -1.77\times 10^{32} \mbox{eV}^{-1}\cong 3.6$ billion years.  That means, the equality of the matter and the scalar field happended at $3.6$ billion years ago. 
And for $w_{eff}\cong -1\Longleftrightarrow \frac{\dot{H}}{H^2}\cong 0$, i.e., the universe enters in a de Sitter phase at late times happens at $N=1.40 $, that means, in terms of the cosmic time, this will happen within $ t-t_0=$  $1.11\times 10^{33} \mbox{eV}^{-1}\cong 22$ billion years.

Finally, it is important to take into account that during the matter domination the dynamical system has an attractor (tracker) solution \cite{pr,pv, liddle}, which satisfies 
\begin{eqnarray}\label{KG}
\ddot{\varphi}+\frac{2}{t}\dot{\varphi}+V_{\varphi}=0.
\end{eqnarray}

In the case of the potential $V_4$, we look for a solution of the form $\varphi_{tra}=-M_{pl}+Ct^{\beta}$ where $C$ and $\beta$ are parameters. Inserting $\varphi_{tra}=-M_{pl}+Ct^{\beta}$ into (\ref{KG}) one obtains
\begin{eqnarray}
\beta(\beta-1)Ct^{\beta-2}+{2\beta}Ct^{\beta-2}-\frac{2m^2M^6}{C^5}t^{-5\beta}=0,
\end{eqnarray}
which is satisfied when $\beta=1/3$ and $C=\left(\frac{9}{2} \right)^{1/6}Mm^{1/3}$, and thus, one gets, $\varphi_{tra}=-M_{pl}+ \left(\frac{9}{2} \right)^{1/6}M(mt)^{1/3}  $.

Taking into account that during matter domination era, one has $t=\frac{2}{3H_0}e^{3N/2}$, we could write
\begin{align}
\varphi_{tra}(N)=-M_{pl}+\left(  \frac{\sqrt{2}m}{H_0} \right)^{1/3}Me^{N/2}\Longrightarrow 
 x_{tra}(N)\equiv \frac{\varphi_{tra}}{M_{pl}}=-1+\left(  \frac{\sqrt{2}m}{H_0} \right)^{1/3}\frac{M}{M_{pl}}e^{N/2} \nonumber \\
 \cong-1+\left( \frac{50\sqrt{2}}{5.94}  \right)^{1/3}\bar{M} e^{N/2}\cong -1+2.28\bar{M} e^{N/2} .
 \end{align}
 
In the same way one can show that
 \begin{eqnarray}
 y_{tra}(N)\equiv \frac{\dot{\varphi}_{tra}}{KM_{pl}}=1.14 \bar{H}_0\bar{M} e^{-N},
 \end{eqnarray} 
 and thus,
\begin{eqnarray}
 \bar{\rho}_{tra}(N)\equiv \frac{y_{tra}^2}{2}+\bar{V}_4(x_{tra})\cong 2.3\times 10^{-3}\bar{M}^2e^{-2N}.  
 \end{eqnarray}
Let us note that the reheating temperature $T_{rh}=100$ TeV which has been considered uniformly throughout the numerical simulation, gives $\bar{M}={7.43}$.
\begin{figure}
\begin{center}
\includegraphics[scale=0.45]{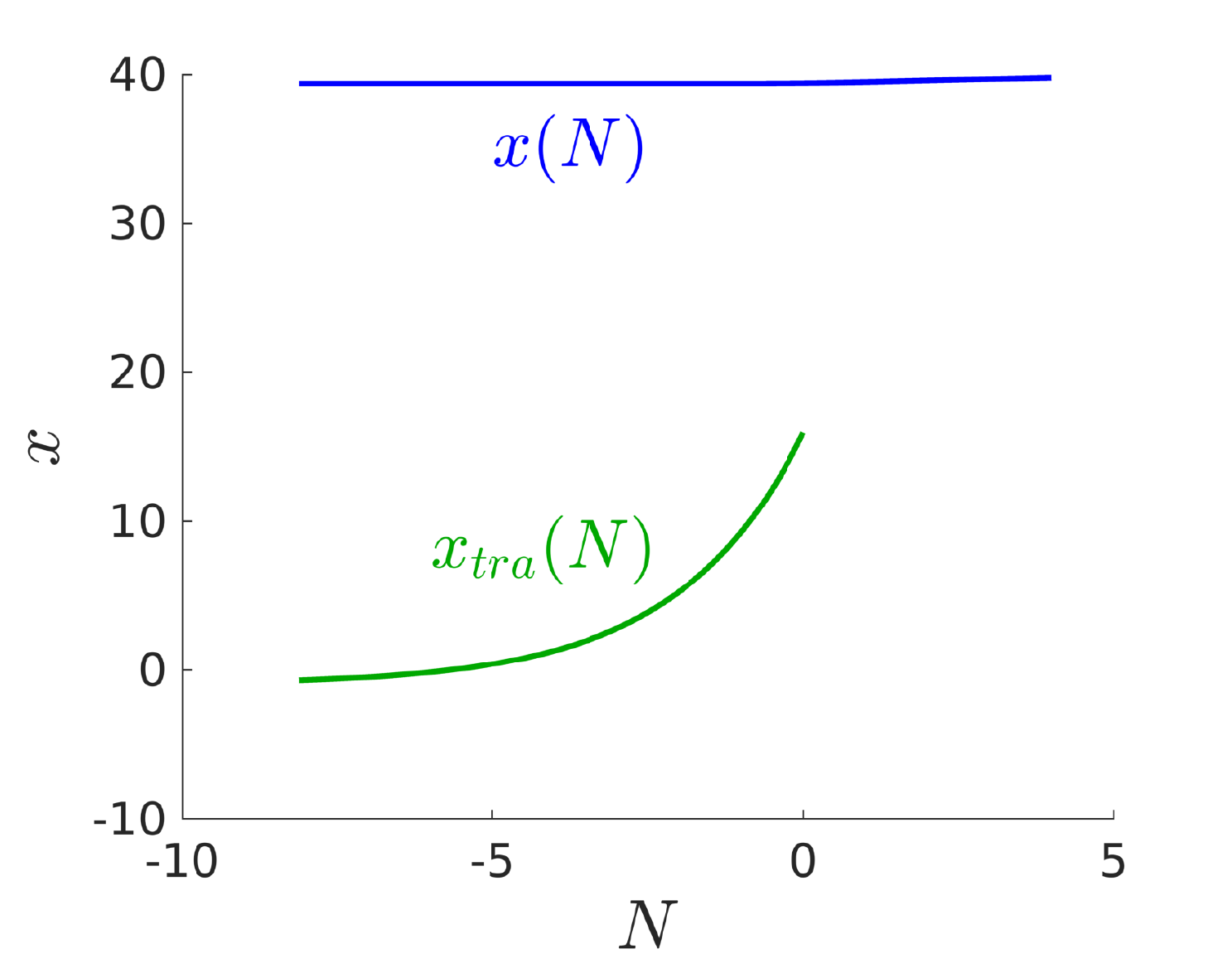}
\end{center}
\caption{(Color online)  Comparison of the dynamics of the tracker solution ($x_{tra}(N)$; green curve) and our solution for the potential $V_4$ ($x(N)$; blue curve) assuming  the reheating temperature $T_{rh} = 100$ TeV. }
\label{fig:modifiedPV3}
\end{figure}
We now close this section with the graphical variations of the cosmological parameters for the quintessence potential. Let us note that in all plots, we have considered the time period from  $N_{eq}$ to $N=4$. In Fig. \ref{fig:modifiedPV1} we describe the evolution of the energy densities of different fluid components in the logarithmic units from which one can see that at present time the scalar field dominates over matter and radiation. The dimensionless density parameters for matter, radiation and the scalar field are shown in the left panel of Fig. \ref{fig:modifiedPV2} in the right panel of  Fig. \ref{fig:modifiedPV2}) we show the evolution of the effective EoS, $w_{eff}$ from which one can clearly visualize a smooth transiton from  $w_{eff} = 0$  to the region with $w_{eff} < -1/3$ and moreover we also find that $w_{eff}$ approaches toward $-1$ in an asymptotic manner. Finally, we compare the evolution of the tracker solution and our solution in Fig. \ref{fig:modifiedPV3} from which we may conclude that similar situation has been depicted in Fig. (2a) of Ref. \cite{dimopoulos1} showing that the tracker solution, which is defined during the matter domination era, does not catch the physical solution $\varphi$ at the present time, and the universe is never driven by the tracker solution.

 \section{Exponential quintessence potential}
\label{sec-exponential}

In this Section we replace the inverse power law quintessence potential of the improved potential given in eqn.  (\ref{improved}) by an exponential one in order to compare with the proposed quintessence potential. Therefore, the quintessential inflation potential in which we are now interested in, takes the following expression
\begin{eqnarray}\label{expo}
V_{\gamma}(\varphi)=\left\{\begin{array}{ccc}
\frac{m^2}{2}(\varphi^2-M_{pl}^2 + M^2) & \mbox{for} & \varphi\leq -M_{pl}\\
\frac{m^2M^2}{2}e^{-\gamma (\frac{\varphi}{M_{pl}}+1)}&\mbox{for} & \varphi\geq -M_{pl},\end{array}
\right.
\end{eqnarray}
where $\gamma$ is a dimensionless parameter. Choosing $K$ as in the previous section and writing $M= \bar{M}\times 10^{-44} M_{pl}$ one can calculate that, 
for $x\geq -1$ one has,
$\bar{V}_{\gamma}(x)=4.14 \times 10^{18}\bar{M}^2 e^{-\gamma(x+1)}$.

Before solving the dynamical system (\ref{system}) using the numerical simulation, first of all, we disregard the radiation component of the energy density because after the matter-radiation equality, the radiation component decreases faster than matter, and following \cite{copeland}, we introduce the dimensionless variables 
\begin{eqnarray}
\tilde{x}\equiv \frac{\dot{\varphi}}{\sqrt{6}M_{pl}H} \quad \mbox{and} \quad \tilde{y}\equiv \frac{\sqrt{V}}{\sqrt{3}M_{pl}H},
\end{eqnarray}
which (after inflation) enable us to write down the following autonomous dynamical system
\begin{eqnarray}
\left\{\begin{array}{ccc}
\tilde{x}' &=&-3\tilde{x}+\sqrt{\frac{3}{2}}\gamma \tilde{y}^2+\frac{3}{2}\tilde{x}\left(\tilde{x}^2-\tilde{y}^2+1\right)\\
\tilde{y}' &=&-\sqrt{\frac{3}{2}}\gamma \tilde{x}\tilde{y}+\frac{3}{2}\tilde{y}\left(\tilde{x}^2-\tilde{y}^2+1\right),\end{array}\right.
\end{eqnarray}
together with the constraint
\begin{eqnarray}
\tilde{x}^2+\tilde{y}^2+\Omega_m=1.
\end{eqnarray}

The interest of the system is that, for $\gamma^2<6$, the point $\left(\tilde{x}=\frac{\gamma}{\sqrt{6}}, \tilde{y}=\sqrt{1-\frac{\gamma^2}{6}}\right)$ is a fixed point of the system. For $\gamma^2<3$ it is a stable node and  for $3<\gamma^2<6$ it is a saddle point \cite{copeland}. Moreover, at the fixed point $\left(\tilde{x}=\frac{\gamma}{\sqrt{6}}, \tilde{y}=\sqrt{1-\frac{\gamma^2}{6}}\right)$, the effective EoS parameter given by
$w_{eff}= \tilde{x}^2-\tilde{y}^2 $, and the density parameter for the quintessence field given by $\Omega_{\varphi}=\tilde{x}^2+\tilde{y}^2 $, respectively take the values, $w_{eff}=\frac{\gamma^2}{3}-1$ and $\Omega_{\varphi}=1$. 
Now, since the late-time acceleration occurs when $w_{eff}<-\frac{1}{3}$, thus, in order to mimic this cosmic acceleration at late time one should have, $\gamma^2<2$. Thus, in order to perform our numerical calculations we will fix a typical value of $\gamma$ satisfying $\gamma^2 <2$. Here, we choose $\gamma=1$,  and integrate the autonomous system (\ref{system}) for the initial conditions obtained when the reheating temperature is $100$ TeV. 
\begin{figure}
\begin{center}
\includegraphics[scale=0.45]{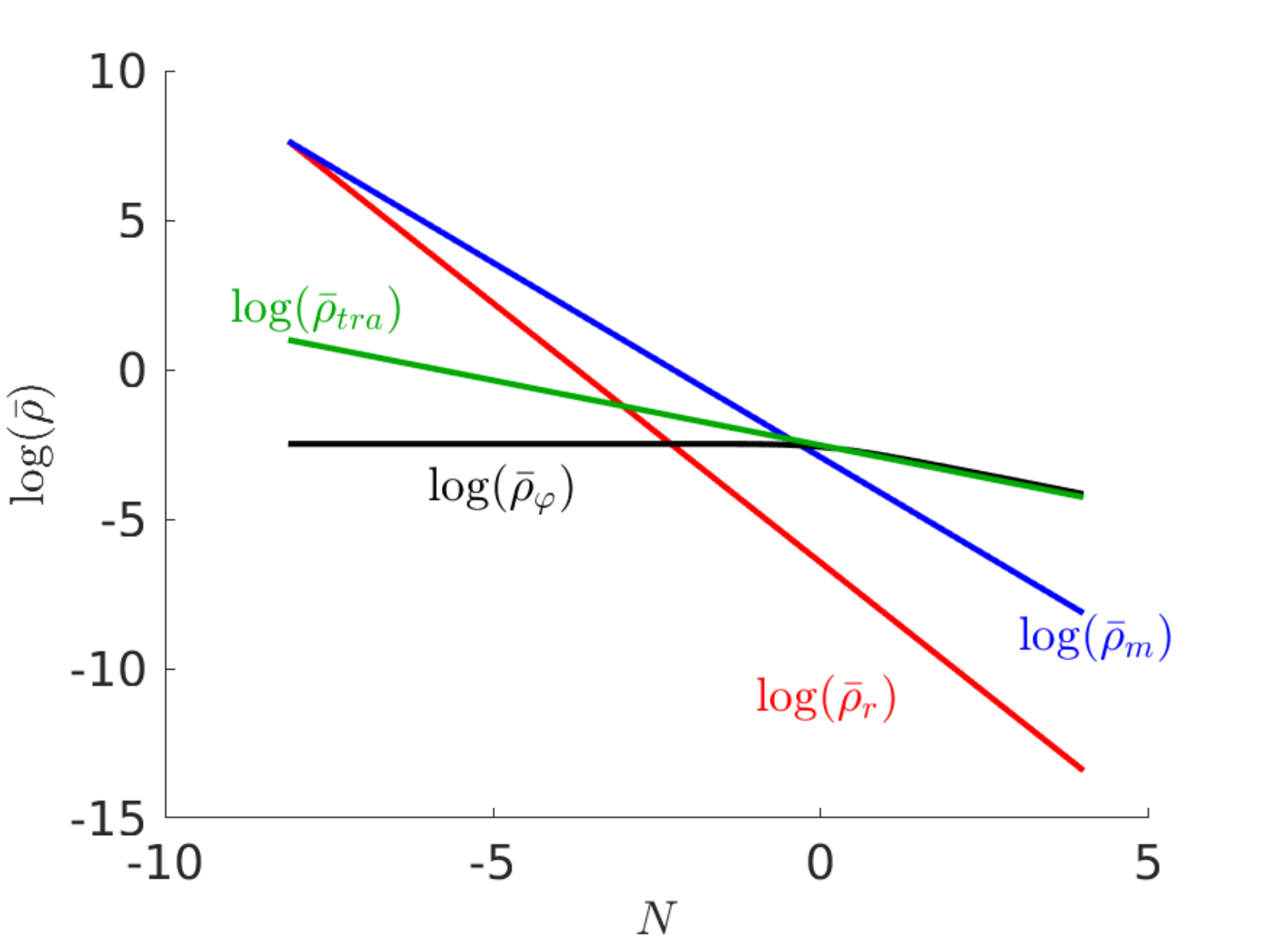}
\end{center}
\caption{(Color online) Evolution of different energy densities has been shown, such as the matter sector (blue curve), radiation (red curve), tracker field (green curve) and the scalar field (black curve) in the logarithmic units. For the numerical simulation we consider $\gamma=1$, reheating temperature $T_{rh}= 100$ TeV and $\bar{M}=0.0167$. The initial conditions are taken at the matter-radiation equality. }
\label{fig:exp1}
\end{figure}
\begin{figure}
\begin{center}
\includegraphics[scale=0.45]{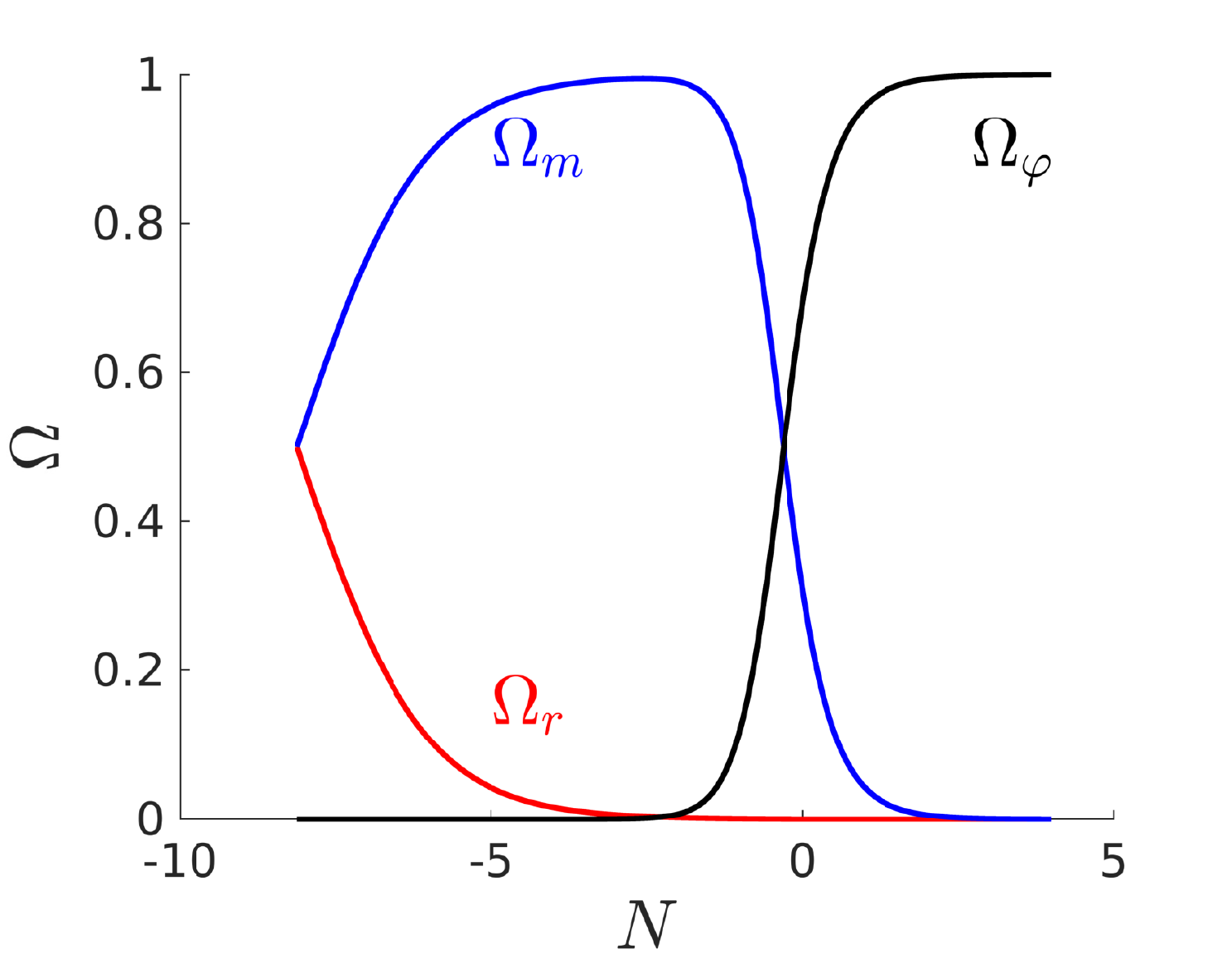}
\includegraphics[scale=0.45]{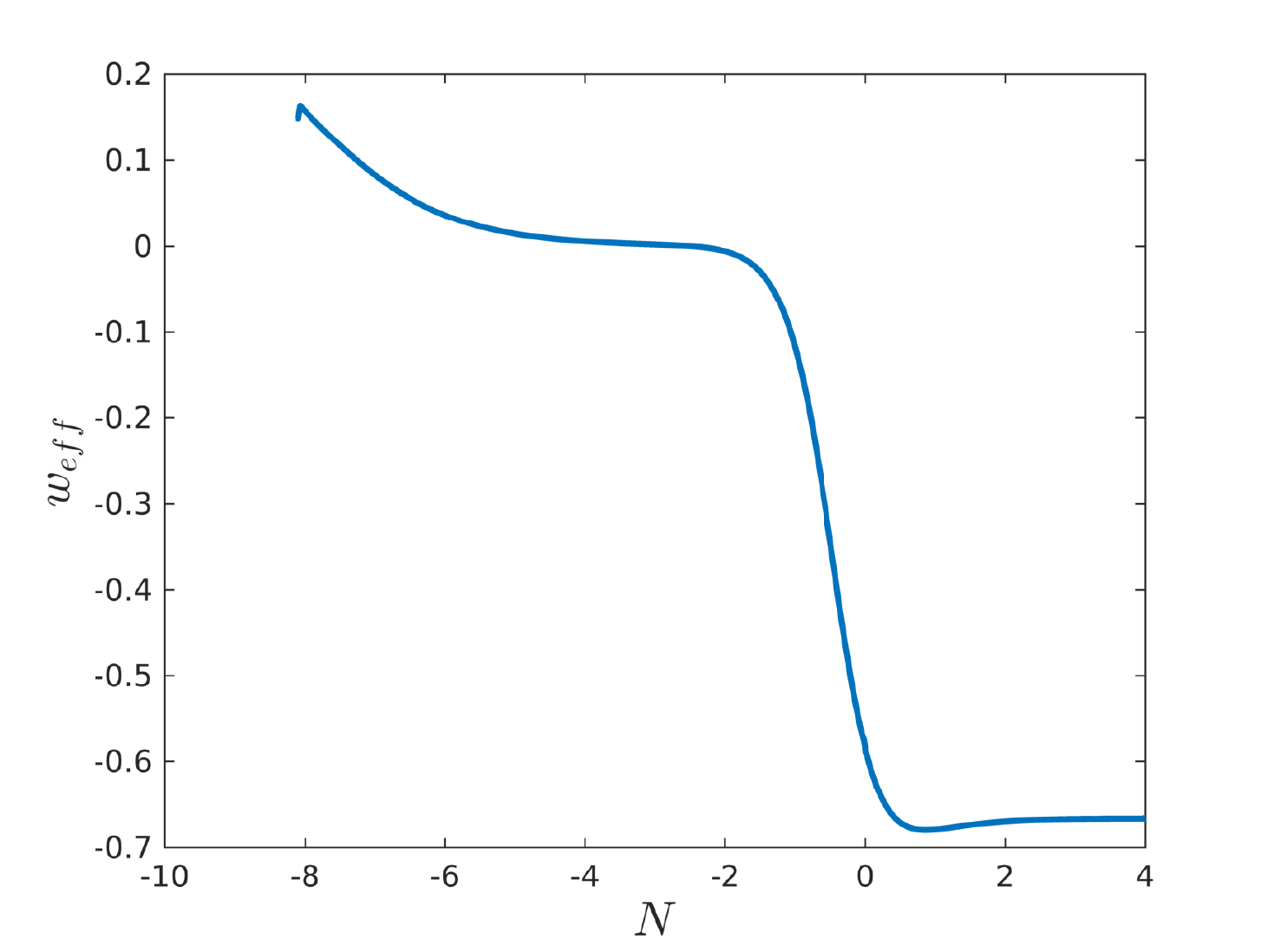}
\end{center}
\caption{(Color online) In the left panel we show the evolution of the density parameter $\Omega$ for matter (blue curve), radiation (red curve) and the scalar field (black curve). In the right panel we present the evolution of the effective EoS  
$w_{eff}$. For the numerical simulation we fix  $\gamma=1$, reheating temperature $T_{rh} = 100$ TeV and $\bar{M}=0.0167$. At the present time $\Omega_{\varphi} \simeq 0.7$, $\Omega_m \simeq 0.3$, $\Omega_{rad} \simeq 0.0$ and  $w_{eff} \simeq -0.6$. 
At late time they tend to $1,0,0$ and $-0.66$ respectively. From the plot of $w_{eff}$ one can see that after the matter-radiation equality the universe enters in a matter domination era, (i.e., $w_{eff}=0$), and finally enters into the current accelerating stage of the universe. }
\label{fig:exp2}
\end{figure}
\begin{figure}
\begin{center}
\includegraphics[scale=0.45]{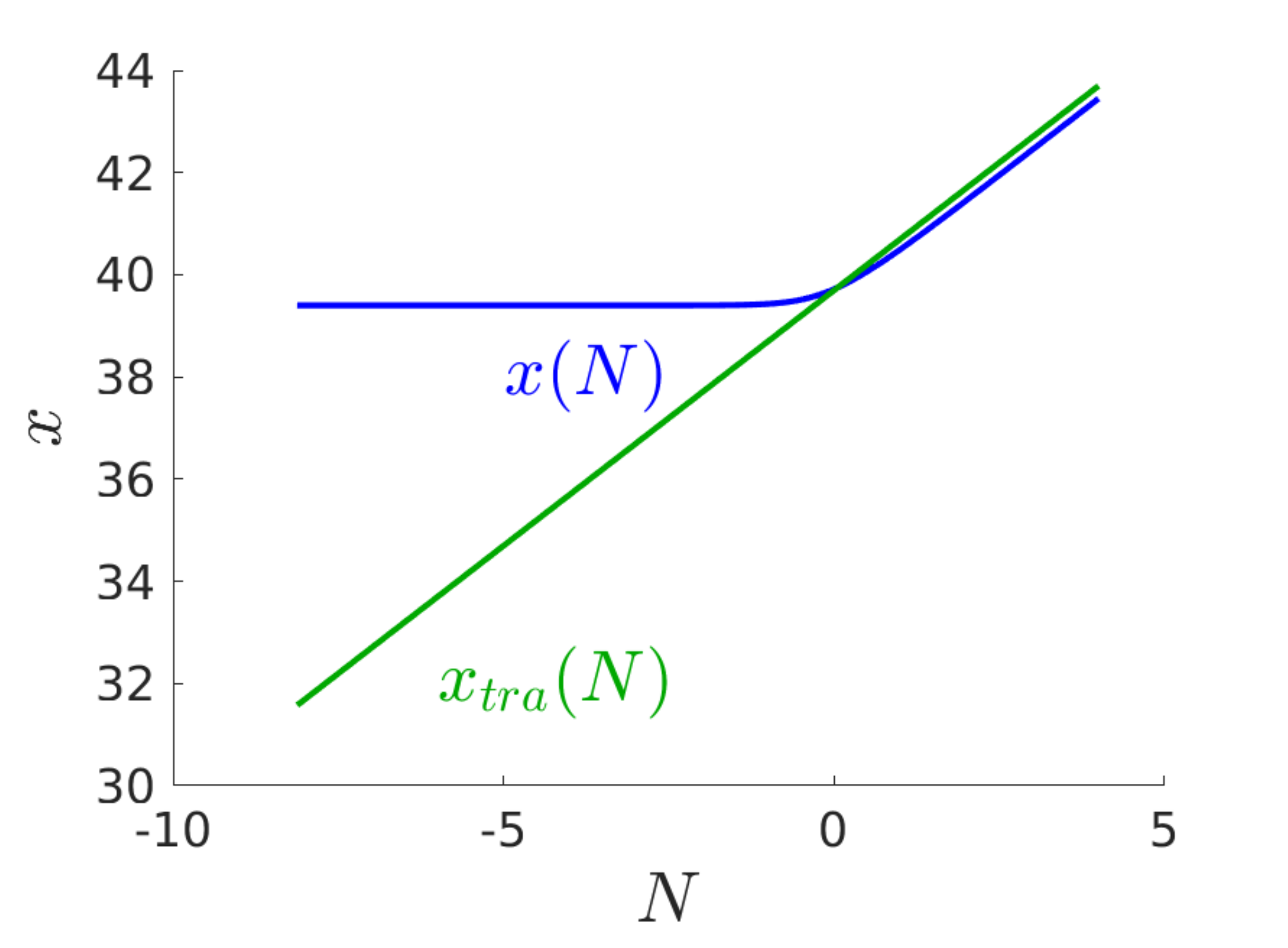}
\end{center}
\caption{Comparison of the dynamics of the tracker solution ($x_{tra}(N)$; green curve) and our solution ($x(N)$; blue curve) for 
the exponential potential  with $\gamma=1$ and reheating temperature $T_{rh} = 100$ TeV.}
\label{fig:exp3}
\end{figure}

Finally, we look for the tracker solution: since for the attractor the scale factor is given by $a\propto t^{\frac{2}{3(w_{eff}+1)}}=t^{2/\gamma^2}$, one has $H=\frac{2}{\gamma^2t}$, and thus, the equation
 $\tilde{x}=\frac{\gamma}{\sqrt{6}}$ leads to
\begin{eqnarray}
\dot{\varphi}_{tra}=\gamma M_{pl}H\Longleftrightarrow \dot{\varphi}_{tra}=\frac{2M_{pl}}{\gamma t}\Longrightarrow \varphi_{tra}(t)=-M_{pl}+\frac{M_{pl}}{\gamma}\ln\left(\frac{t^2}{\bar{t}^2} \right).
\end{eqnarray}

To obtain the value of the parameter $\bar{t}$ we use the equation $\tilde{y}=\sqrt{1-\frac{\gamma^2}{6}}$, getting $\bar{t}^2=\frac{4(6-\gamma^2)M_{pl}^2}{\gamma^4m^2M^2}$. Then, the tracker solution is
\begin{eqnarray}
\varphi_{tra}(t)=-M_{pl}+\frac{M_{pl}}{\gamma}\ln\left(\frac{\gamma^4m^2M^2}{4(6-\gamma^2)M^2_{pl}}{t^2} \right).
\end{eqnarray}

To write it as a function of the time $N=-\ln(1+z)=\ln\left(\frac{a}{a_0} \right)$, we use that for the EoS parameter $w_{eff}=\frac{\gamma^2}{3}-1$, one has
\begin{eqnarray}
\frac{\rho(t)}{\rho_0}=\left(\frac{a_0}{a} \right)^{3(w_{eff}+1)}=\left(\frac{a_0}{a} \right)^{\gamma^2}
\Longrightarrow H=e^{-\gamma^2N/2}H_0\Longrightarrow t=\frac{2}{\gamma^2H_0}e^{\gamma^2N/2},\end{eqnarray}
and then,
\begin{eqnarray}
\varphi_{tra}(N)=-M_{pl}+\gamma M_{pl}N+\frac{M_{pl}}{\gamma}\ln\left( \frac{m^2M^2}{(6-\gamma^2)M_{pl}^2H_0^2}    \right)\nonumber \\
\Longrightarrow x_{tra}(N)=-1+\gamma N+\frac{1}{\gamma}\ln\left( \frac{m^2M^2}{(6-\gamma^2)M_{pl}^2H_0^2}    \right).
\end{eqnarray}

In the same way,
\begin{eqnarray}
\dot{\varphi}_{tra}(t)=\frac{2M_{pl}}{\gamma t}\Longrightarrow y_{tra}(N)=\gamma \bar{H}_0e^{-\gamma^2 N/2}.
\end{eqnarray}

And thus, its energy density is given by,
\begin{eqnarray}
\bar{\rho}_{tra}(N)=\frac{y_{tra}^2}{2}+V_{\gamma}(x_{tra})=
3\bar{H}_0^2 e^{-\gamma^2N}.
\end{eqnarray}

Now we close this section by presenting the graphical variations of the cosmological parameters for this quintessence potential.  In Fig. \ref{fig:exp1}, Fig. \ref{fig:exp2}, and Fig. \ref{fig:exp3} we show the graphical variations of various cosmological parameters. In particular, in Fig. \ref{fig:exp1} we describe the evolution of different energy densities in the logarithmic units which exhibit the similar behavior as described in  Fig. (2d) of \cite{dimopoulos1}.  In Fig. \ref{fig:exp2} we show the evolution of the density parameters (left graph of Fig. \ref{fig:exp2}) and the evolution of the effective EoS (right graph of Fig. \ref{fig:exp2}). Finally, in Fig. \ref{fig:exp3} we compare the tracker solution and our solution for the exponential potential from which one can clearly see that the tracker field  catches the scalar field $\varphi$ at the present time.

\section{Concluding remarks}
\label{sec-summary}

The  quintessential inflation model by Peebles and Vilenkin (PV) is an elegant unified description for the early- and late- evolutions of the universe \cite{pv} and this is the first attempt to unify these distant phases using a single scalar field potential. For extreme simplicity of the model and its potentiality as well, the PV model certainly gained a massive attention to the cosmological community since the end of nineties. Subsequently, with the rapid developments in the observational science, the theory of quintessential inflation has become a major area of cosmology for further examinations with a hope to offer a observationally viable single theoretical description for the universe's evolution starting from its early phase to current stage.

Although the PV model is quite classic connecting inflation to quintessence, but according to the observational data, the model needs some simple improvements for the following reasons. For the inflationary piece of the model described by the quartic potential, the theoretical values of the spectral index ($n_s$) and the ratio of tensor to scalar perturbations ($r$) do not enter into the corresponding two-dimensional marginalized joint contour at $95\%$ CL as reported by \cite{Planck}. However, on the contrary, if the quartic part of the inflationary model is replaced by the quadratic function of the potential,  then this problem does not appear. Additionally, for the reheating mechanism adopted in \cite{pv}, the gravitational particle production of massless particles gives a reheating temperature of the order of $1$ TeV, which according to the observational predictions is not so able to solve the overproduction of Gravitational Waves (GW). As a consequence, this might affect the success of the Big Bang Nucleosynthesis process.

Thus, keeping the above limitations, we have taken an attempt to perform some simple modifications of the PV model in agreement with the observational bounds. We have replaced the quartic piece of the inflationary potential by the quadratic one and consider the gravitational production of heavy massive particles for the reheating of our universe \cite{haro18}. We find that the newly constructed quintessential inflation model, whose parameter $M$ depends on the reheating temperature, can well behave with the observational data. In fact, we have studied the evolution of the universe starting from the matter-radiation equality up to present time by presenting the graphical variations of various cosmological parameters. In terms of the effective equation of state, $w_{eff}$, we find that after the matter-radiation equality, the universe enters into the matter dominated era ($\equiv w_{eff} = 0$) and then to the present accelerating epxansion ($w_{eff} < -1/3$) and consequently, $w_{eff} \rightarrow -1$, in an asymptotic manner.  Finally, in section \ref{sec-exponential}, we have made a comparison of the quintessence potential of the improved PV model to that of an exponential potential.

Last but not least, we anticipate that the proposed improved version of the quintessential inflationary model will  offer a new range of possibilities, both from theoretical and observational grounds because the  improvements that we propose should subsequently improve other cosmological parameters in a similar fashion and this will enable one to understand the real improvement of this model. From the observational direction, in particular, there are several interesting investigations can be performed. The constraints using likelihood from Planck 2018 is a necessary work. Moreover, the inclusion of next generation of cosmological data sets will be surely interesting in order to quantify the free parameters of this model for a better conclusion. These all are kept for future works.

\vspace{1cm}

{\it Acknowledgments.}   
This investigation has been supported by MINECO (Spain) grants  MTM2014-52402-C3-1-P and MTM2017-84214-C2-1-P, and  in part by the Catalan Government 2017-SGR-247. SP acknowledges the research grant through the Faculty Research and Professional Development Fund (FRPDF) Scheme of Presidency University, Kolkata, India.

\end{document}